\documentclass[journal]{IEEEtran}

\ifCLASSINFOpdf
  \usepackage[pdftex]{graphicx}
  \DeclareGraphicsExtensions{.pdf,.jpeg,.png}
\else
  \usepackage[dvips]{graphicx}
\fi
\usepackage{epstopdf}
\usepackage[cmex10]{amsmath}
\usepackage{amssymb}
\usepackage{wasysym}
\usepackage{algorithm}
\usepackage{algorithmic}
\usepackage{array}
\usepackage{cite}
\usepackage{color}
\usepackage{url}

\usepackage{epsfig,latexsym}
\usepackage{flushend}
\usepackage{cite}
\usepackage{verbatim}
\usepackage{amsopn}
\usepackage{booktabs}

\usepackage{hyperref}

\begin{document}

\title{Machine Learning for Beam Alignment in Millimeter Wave Massive MIMO}


\author{Wenyan~Ma,~\IEEEmembership{Student~Member,~IEEE}, Chenhao~Qi,~\IEEEmembership{Senior~Member,~IEEE}, \\ and Geoffrey Ye Li,~\IEEEmembership{Fellow,~IEEE}
\thanks{This work was supported in part by National Natural Science Foundation of China under Grant 61871119, by Natural Science Foundation of Jiangsu Province under Grant BK20161428, and by the Fundamental Research Funds for the Central Universities. (\textit{Corresponding author: Chenhao~Qi})}
\thanks{Wenyan~Ma and Chenhao~Qi are with the School of Information Science and Engineering, Southeast University, Nanjing 210096, China (Email: qch@seu.edu.cn).}
\thanks{Geoffrey Ye Li is with the School of Electrical and Computer Engineering, Georgia Institute of Technology, Atlanta, GA, USA (Email: liye@ece.gatech.edu).}
}

\markboth{Accepted By IEEE Wireless Communications Letters}
{Shell \MakeLowercase{\textit{et al.}}: Bare Demo of IEEEtran.cls for Journals}

\maketitle

\begin{abstract}
This  article investigates beam alignment for multi-user millimeter wave (mmWave) massive multi-input multi-output system. Unlike the existing works using machine learning (ML), an alignment method with partial beams using ML (AMPBML) is proposed without any prior knowledge such as user location information. The neural network (NN) for the AMPBML is trained offline using simulated environments according to the mmWave channel model and is then deployed online to predict the beam distribution vector using partial beams. Afterwards, the beams for all users are all aligned simultaneously based on the indices of the dominant entries of the obtained beam distribution vector. Simulation results demonstrate that the AMPBML  outperforms the existing methods, including the adaptive compressed sensing, hierarchical search, and multi-path decomposition and recovery, in terms of the total training time slots and the spectral efficiency.

\end{abstract}
\begin{IEEEkeywords}
Beam alignment, machine learning, massive MIMO, millimeter wave communications.
\end{IEEEkeywords}

\section{Introduction}
Due to its abundant frequency spectrum resource, millimeter wave (mmWave) communications have attracted broad attention and become an important technology in the future~\cite{alkhateeb2014channel}. The mmWave signal experiences high path loss but can be compensated by utilizing a massive  multi-input multi-output (MIMO) antenna array to achieve directional beam alignment (BA) and data transmission. Its short wavelength enables large antenna arrays to be packed into small form factors.

In multi-user multi-stream mmWave systems, the base station (BS) simultaneously serves multiple users with multiple beams. To align beams for different users, hierarchical codebook based beam training is usually used.  For examples, an adaptive compressed sensing (ACS)  method has been proposed in~\cite{alkhateeb2014channel}, where a  hierarchical codebook with multi-resolution is designed to train beams for all users sequentially. Then a hierarchical search (HS) method, which can be performed much faster than the ACS  method, has been developed in~\cite{xiao2016hi}. To combine the advantages of the HS and the ACS, a multi-path decomposition and recovery (MDR) method has been proposed in~\cite{xiao2018enhanced} . However, using hierarchical codebook based beam training to align beams for multiple users is not trivial. The BS has to align beams for all users sequentially during the training stage, leading to huge overhead. Moreover, the optimal codeword index must be fed back for each layer of the hierarchical codebook, which is also time consuming. Other work investigates the beam alignment in mmWave systems equipped with automotive sensors or radars, where the BS can obtain the user location information from these sensors or radars and the directional beams can be designed.  For examples, a beam alignment solution is designed by extracting useful information from radar signal~\cite{gon2016radar}, while the beams are aligned based on the location information from automotive sensors~\cite{choi2016millimeter}. However, the equipped automotive sensors or radars will incur additional hardware overhead.

Recently, machine learning (ML) has been applied to address different issues in physical layer communications~\cite{qin2019deep}. The application of ML to BA in mmWave systems has also been investigated to take the advantages of ML in solving complicated  nonlinear problems. For examples, the ML tools and situational awareness are combined to learn the beam information including power and optimal beam index~\cite{wang2018mmwave} while the angles of arrival (AoAs) can be estimated and input to the neural network (NN) for beam selection~\cite{anton2019learning}. However, the above two ML methods need the location information of the users to train the NN, which incurs extra system overhead.

In this article, we propose an alignment method with partial beams using ML (AMPBML) for the multi-user mmWave massive MIMO system. The NN for the AMPBML is trained offline using simulated environments according to the mmWave channel model and is then deployed online to predict the beam distribution vector using partial beams. Afterwards, the beams for all users are aligned simultaneously based on the obtained indices of the dominant entries of beam distribution vector. Different from the existing works based on hierarchical codebook, we align beams for all users simultaneously and significantly save the total training time. Moreover, unlike the existing works on BA using ML, we need no prior knowledge, such as the user location information, to train the NN. It will reduce the system overhead significantly.

\textit{Notations:} Symbols for  vectors (lower case) and matrices (upper case) are in boldface. $(\cdot)^T $, $(\cdot)^* $,  $(\cdot)^H $, and $(\cdot)^{-1} $ denote the transpose, conjugate,  conjugate transpose, and inverse, respectively. We use $\boldsymbol{I}_{L}$ to represent identity matrix of size $L$. The set of $M\times{N}$ complex-valued matrices and integral-valued matrices is denoted as $\mathbb{C}^{M\times{N}}$ and $\mathbb{Z}^{M\times{N}}$, respectively. We use $\mathbb{E}\{\cdot\}$ to denote expectation. $\|\cdot\|_2$ and $\|\cdot\|_F$ denote $l_2$-norm of a vector and Frobenius norm of a matrix, respectively. The $n$th entry of $\boldsymbol{a}$ is denoted as $\boldsymbol{a}[n]$. We use $\lfloor \cdot \rfloor$ and $\lceil \cdot \rceil$ to denote floor and ceil operations, respectively. Complex Gaussian distribution is denoted as $\mathcal{CN}$.

\section{System Model and Problem Formulation}
We first introduce the multi-user mmWave massive MIMO system. Then we formulate the BA problem for multiple users.

\subsection{System Model}
We consider an uplink multi-user mmWave massive MIMO communication system comprising a BS and $U$ users. The BS is equipped with a uniform linear array (ULA)~\cite{alkhateeb2014channel}. Note that the present method can be generalized to other array structures. Hybrid combining is typically adopted, where the number of antennas, $N_A$, is much larger than that of RF chains, $N_R$, i.e., $  N_A \gg N_R$.

For uplink transmission, hybrid combining at the BS consists of baseband digital combining and RF analog combining~\cite{sun2019beam}. Denote $s_u$ to be the transmit signal. Then the received signal vector at the BS can be expressed as
\begin{equation}\label{UpSig}
\boldsymbol{y} = \boldsymbol{W}_B^H \boldsymbol{W}_R^H  \sum_{u=1}^{U} \boldsymbol{h}_u s_u + \boldsymbol{W}_B^H \boldsymbol{W}_R^H \boldsymbol{n},
\end{equation}
where $\boldsymbol{W}_B\in{\mathbb{C}^{N_R\times{N_R}}}$ and $\boldsymbol{W}_R\in{\mathbb{C}^{N_A\times{N_R}}}$ are the digital combining matrix and analog combining matrix, respectively, and $\boldsymbol{n} \in{\mathbb{C}^{N_A}} $ is the additive white Gaussian noise (AWGN) vector satisfying $\boldsymbol{n} \sim \mathcal{CN}(0,\sigma^{2}\boldsymbol{I}_{N_A})$. To normalize the power of the hybrid combiner, we set $\| \boldsymbol{W}_B^H \boldsymbol{W}_R^H \| _F ^2= N_R$ and $\mathbb{E}\{s_u{s_u}^*\}=1$.

There are different kinds of channel model in mmWave systems, such as the clustered mmWave channel model~\cite{ni2017near,yan2018tracking} and the Saleh-Valenzuela mmWave channel model~\cite{alkhateeb2014channel}. We choose the Saleh-Valenzuela mmWave channel model in our paper. For the Saleh-Valenzuela mmWave channel
model~\cite{alkhateeb2014channel}, the channel vector $\boldsymbol{h}_u\in{\mathbb{C}^{N_A}}$ between the $u$th user and the BS can be represented by
\begin{equation}\label{hu}
\boldsymbol{h}_{u}=\sqrt{\frac{N_A}{L_u}}\sum_{i=1}^{L_u} \boldsymbol{h}_{u,i}=\sqrt{\frac{N_A}{L_u}}\sum_{i=1}^{L_u} g_{u,i} \boldsymbol{\alpha} (N_A,\theta_{u,i}),
\end{equation}
where $\boldsymbol{h}_{u,i}$, $L_{u}$,  and $g_{u,i}$ are denoted as the channel vector, number of multiple channel paths,  and complex gain of the $i$th path, respectively, and the steering vector $\boldsymbol{\alpha}(N,\theta)$ in \eqref{hu} can be expressed as $\boldsymbol{\alpha}(N,\theta)= \frac{1}{\sqrt{N}}\left[1,e^{j\pi\theta},...,e^{j\pi\theta(N-1)}\right]^{T}$~\cite{xiao2016hi,xiao2018enhanced}. Typically $\boldsymbol{h}_{u}$ consists of one  line-of-sight (LOS) path (the 1st channel path), and $L_u-1$ non-line-of-sight (NLOS) paths (the $i$th channel path for $2\leq i \leq L_u)$.

The AoA of the $i$th path of the $u$th user $\vartheta_{u,i}$ is uniformly distributed over $[-\pi,\pi)$ \cite{alkhateeb2014channel}. Then $\theta_{u,i} \triangleq \sin{\vartheta_{u,i}}$ in  (\ref{hu})  if the distance between adjacent antennas at the BS is with half-wave length.

\subsection{Problem Formulation}
According to~\cite{sun2019beam}, each column of the analog combiner is a codeword selected from a beam steering codebook, which consists of $N_A$ equally spaced channel steering vectors pointing at $N_A$ different directions. The codebook at the BS is denoted by $\boldsymbol{\mathcal{W}} \triangleq \{ \boldsymbol{w}(1), \boldsymbol{w}(2), \ldots, \boldsymbol{w}(N_A) \}$, where
\begin{equation}
\boldsymbol{w}(n) = \boldsymbol{\alpha}(N_A,-1+(2n-1)/N_A).
\end{equation}
Our objective is to search $U$ beams from the total $N_A$ ones that align with the LOS paths of $U$ users. Denote $\boldsymbol{C} \triangleq [ \boldsymbol{w}(1), \boldsymbol{w}(2), \ldots, \boldsymbol{w}(N_A) ] \in{\mathbb{C}^{N_A\times{N_A}}}$ and $\boldsymbol{\beta}_u \triangleq \boldsymbol{C}^H \boldsymbol{\alpha}(N_A,\theta_{u,1}) \in{\mathbb{C}^{N_A}}$. Then the objective can be denoted as
\begin{equation}\label{walpha}
\arg\underset{n=1,2,\ldots,N_A} {\max} \Big| \boldsymbol{\beta}_u[n] \Big|, ~~u=1,2,\ldots,U.
\end{equation}
Denote $\boldsymbol{b} \triangleq [b_1, b_2, \ldots, b_u]^T \in{\mathbb{Z}^{U}}$ as the $U$ beam indices that align with the LOS paths of $U$ users, where $b_u$ can be denoted as
\begin{equation}\label{bu}
b_u = \left\lfloor \frac{N_A( \theta_{u,1}+1)}{2} \right\rfloor +1 \in \{ 1,2,\ldots,N_A \}.
\end{equation}

Given $\theta_{u,1}$, the objective $U$ beam indices can be directly calculated according to \eqref{bu}. However, $\theta_{u,1}$ is unknown before channel training. Exhaustive search to solve \eqref{walpha} for the best analog combiner requires $N_A U / N_R$ time slots. For example, $N_A U / N_R = 256$ if $N_A=256$, $U=3$, and $N_R=3$~\cite{sun2019beam}. The exhaustive search requiring the full knowledge of $\boldsymbol{\beta}_u$ is rather time consuming if there are a large number of antennas. In order to reduce the time slots, we use only part of total $N_A$ beams to predict the optimal $U$ beams. However, due to the nonlinearity of $\boldsymbol{\beta}_u$, it is hard to directly calculate the optimal beam based on partial entries of $\boldsymbol{\beta}_u$. Recently, ML has shown advantages on dealing with nonlinear problems. Therefore we will develop ML based BA based on partial entries of $\boldsymbol{\beta}_u$.

\section{AMPBML}
The major steps of  the AMPBML is summarized in Algorithm~\ref{alg1}. During the uplink beam training, we use $J$ different analog combining matrices, denoted as $\boldsymbol{W}_R^t\in{\mathbb{C}^{N_A\times{N_R}}}$. The beam alignment chooses analog beams that best align with the LOS path of each user, which is independent of the digital combining matrix. Therefore we set $\boldsymbol{W}_B=\boldsymbol{I}_{N_R}$ at the BS. For simplicity, each user transmits the same signal $s_u=1$ for all $J$ slots. The channel is assumed to be time-invariant during $J$ time slots. The received signal vector at the BS at the $t$th slot can be denoted as
\begin{equation}
\boldsymbol{y}^t = (\boldsymbol{W}_R^t)^H  \sum_{u=1}^{U} \boldsymbol{h}_u  +  \tilde{\boldsymbol{n}}^t,
\end{equation}
where $\tilde{\boldsymbol{n}}^t \triangleq (\boldsymbol{W}_R^t)^H \boldsymbol{n}^t \in{\mathbb{C}^{N_R}}$ and $\boldsymbol{n}^t \in{\mathbb{C}^{N_A}}$ is the AWGN vector, each entry of which is with independent complex Gaussian distribution with zero mean and variance of $\sigma^2$.

Denote $M \triangleq J N_R$. We stack the $J$ received signal vectors together and have
\begin{equation}\label{r}
\boldsymbol{r} = (\boldsymbol{W})^H  \sum_{u=1}^{U} \boldsymbol{h}_u  +  \tilde{\boldsymbol{n}},
\end{equation}
where
\begin{equation}\label{rwn}
\begin{split}
 \boldsymbol{r} &\triangleq [(\boldsymbol{y}^1)^T , (\boldsymbol{y}^2)^T , \ldots , (\boldsymbol{y}^J)^T]^T \in{\mathbb{C}^{M}}, \\
 \boldsymbol{W} &\triangleq [\boldsymbol{W}_R^1 , \boldsymbol{W}_R^2 , \ldots , \boldsymbol{W}_R^J] \in{\mathbb{C}^{N_A \times{M}}}, \\
 \tilde{\boldsymbol{n}} &\triangleq [(\tilde{\boldsymbol{n}}^1)^T , (\tilde{\boldsymbol{n}}^2)^T , \ldots , (\tilde{\boldsymbol{n}}^J)^T]^T \in{\mathbb{C}^{M}}.
\end{split}
\end{equation}
To obtain partial entries of $\boldsymbol{\beta}_u$, $\boldsymbol{W}$ is set as a submatrix consisted of $M$ columns of $\boldsymbol{C}$. Since $\boldsymbol{r}$ is determined by $\boldsymbol{h}_u$ as shown in \eqref{r} while $\boldsymbol{h}_u$ is related to the AoA of the LOS path $\vartheta_{u,1}$ as shown in \eqref{hu}, $\boldsymbol{r}$ contains the information of the AoA of the LOS path, indicating that we do not require other information except $\boldsymbol{r}$ to train the NN.

\begin{figure}[!t]
\centering
\includegraphics[width=90mm]{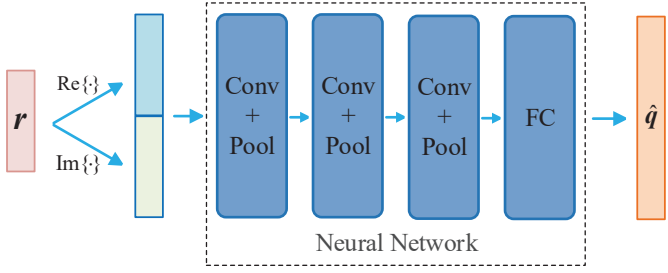}
\caption{Illustration of the NN.}
\label{FIG3}
\end{figure}

\begin{algorithm}[!t]
	\caption{  AMPBML}
	\label{alg1}
	\begin{algorithmic}[1]
		\STATE \emph{Input:} $\boldsymbol{y}^t$, $J$.

        \STATE Obtain $\boldsymbol{r}$ via (\ref{rwn}).
        \STATE Input $\boldsymbol{r}$ to the  offline-trained NN to get $\hat{\boldsymbol{q}}$.
        \STATE Obtain $\boldsymbol{v}$ by sorting the absolute value of $\hat{\boldsymbol{q}}$ in descending order.
        \STATE Calculate $\hat{\boldsymbol{b}}$ based on (\ref{hatb}).

        \STATE \emph{Output:} $\hat{\boldsymbol{b}}$.
	\end{algorithmic}
\end{algorithm}

The BA for multiple users has two stages, the offline training of the NN and online deployment. The NN is first trained offline and then used to predict the $U$ beam indices. As in Fig.~\ref{FIG3}, the input of the NN is $\boldsymbol{r}$. The beam distribution vector, denoted by $\boldsymbol{q} \in{\mathbb{Z}^{N_A}}$, can be represented as
\begin{equation}\label{q}
\boldsymbol{q}[n] = \left\{
\begin{aligned}
1, ~~& n = b_u, u=1,2,\ldots,U, \\
0, ~~& \rm{otherwise}.\\
\end{aligned}
\right.
\end{equation}
The indices of the nonzero entries of $\boldsymbol{q}$ represent the targeted $U$ beam indices.  Note that the AoA is known only during the offline training stage while it is unknown during
the online deployment stage.  The output of the NN is denoted by $\hat{\boldsymbol{q}}$ and is expected to be as close to $\boldsymbol{q}$ as possible.

As illustrated in Fig.~\ref{FIG3}, the adopted NN in this work consists of three hidden layers and a fully connected (FC) layer. Since the NN can only deal with the real number, the input of the NN is a real-valued vector with the length of $2M$ composed by the imaginary and real  parts of $\boldsymbol{r}$.  Each hidden layer consists of a convolutional (Conv) layer and a pooling (Pool) layer. We use two NN parameter sets: NN parameter set (NPS) I and NPS II. The strides of each Conv layer is set to be 1. The kernel size of each Conv layer is set to be 5  and 10 in NPS I and NPS II, respectively. The numbers of filters of these three Conv layers are set as 16, 32, and 64, respectively, in NPS I and 32, 64, and 128, respectively, in NPS II. The activation function of the Conv layer is the ReLU function, that is $f_{\rm Re}(x)=\max(0,x)$. Both the pool size and strides of each Pool layer are set to be 2.

During the offline training of the NN, we generate the dataset of $\boldsymbol{r}$ and $\boldsymbol{q}$ based on the simulated mmWave channel environment. With the beam distribution vector in \eqref{q} and the received signals in \eqref{r}, the training data of $\boldsymbol{r}$ and $\boldsymbol{q}$ can be obtained. In fact, the process to obtain $\boldsymbol{r}$ and $\boldsymbol{q}$  involves the following four steps.

\textbf{i)} Randomly generate a channel vector based on the mmWave channel model in \eqref{hu};

\textbf{ii)} obtain $b_u$ based on \eqref{bu};

\textbf{iii)} enumerate the beam distribution vector $\boldsymbol{q}$ based on \eqref{q};

\textbf{iv)} compute the received signal vector $\boldsymbol{r}$ based on \eqref{r}.
\\ We divide the data set into the training set and the validation set randomly, where the size of the training set is nine times the size of the validation set. The output of the NN is $\hat{\boldsymbol{q}}$.

The training of the NN aims to minimize the difference between $\hat{\boldsymbol{q}}$ and $\boldsymbol{q}$, called the loss in ML. It can be calculated in several
ways. In our work, we formulate the beam alignment problem as a multi-label classification  problem and  use the cross-entropy loss with the sigmoid function as~\cite{wang2018mmwave}
\begin{equation}\label{floss}
f_{\rm Loss}(\boldsymbol{q}, \hat{\boldsymbol{q}}) = \frac{1}{N_A} \sum_{n=1}^{N_A}   \max(\hat{\boldsymbol{q}}[n], 0) - \hat{\boldsymbol{q}}[n] \boldsymbol{q}[n] + \ln(1+e^{-|\hat{\boldsymbol{q}}[n]|}).
\end{equation}

The adaptive moment estimation (Adam) optimizer is used to train the NN by TensorFlow. The NN is trained
for 6,000 epochs, where 500 mini-batches are utilized in each epoch. The learning rate is set to be a step function and decreases with the increasing of training epochs. The learning rate is initialized with the  value of 0.01 and decreases 5-fold every 1,000 epochs.

During the online deployment of the NN, we obtain the real measured $\boldsymbol{r}$ from practical mmWave channel environments, which is then input to the offline-trained NN. The prediction of $\boldsymbol{q}$ by the NN is $\hat{\boldsymbol{q}}$. Due to the mismatching channel environment between the offline training stage and the online deployment stage, $\hat{\boldsymbol{q}}$ may not have only $U$ nonzero entries. Therefore, we select $U$ dominant entries of $\hat{\boldsymbol{q}}$ as  the prediction of the $U$ beam indices that align with the LOS paths of $U$ users. Denote $\hat{\boldsymbol{b}} \triangleq [\hat{b}_1, \hat{b}_2, \ldots, \hat{b}_U]^T \in{\mathbb{Z}^{U}}$ as the prediction of the $U$ beam indices, which can be represented as
\begin{equation}\label{hatb}
\hat{b}_u = \boldsymbol{v}[u],
\end{equation}
for $u=1,2,\ldots,U$.

\section{Simulation Results}
We consider an mmWave massive MIMO system with a BS equipped with $N_A=256$ antennas and $N_R=8$ RF chains. The number of resolvable paths in the mmWave channel is set to be $L_{u}=3$, while $g_{u,1}\thicksim\mathcal{CN}(0,1)$ and $g_{u,i}\thicksim\mathcal{CN}(0,0.01)$ for $i=2,3$~\cite{sun2019beam}. We use $J=16$ time slots to transmit signals for uplink channel estimation. Then we have $M=J N_R=128$. $\boldsymbol{W}$ is set as a submatrix consisted of $M$ columns from $\boldsymbol{C}$, with column indices denoted as $1,1+\lfloor N_A/M \rfloor ,\ldots,1+(M-1) \lfloor N_A/M \rfloor$ from $\boldsymbol{C}$. We compare the proposed AMPBML with the ACS~\cite{alkhateeb2014channel}, HS~\cite{xiao2016hi}, and MDR~\cite{xiao2018enhanced} beam training methods.

\begin{figure}[!t]
\centering
\includegraphics[width=90mm]{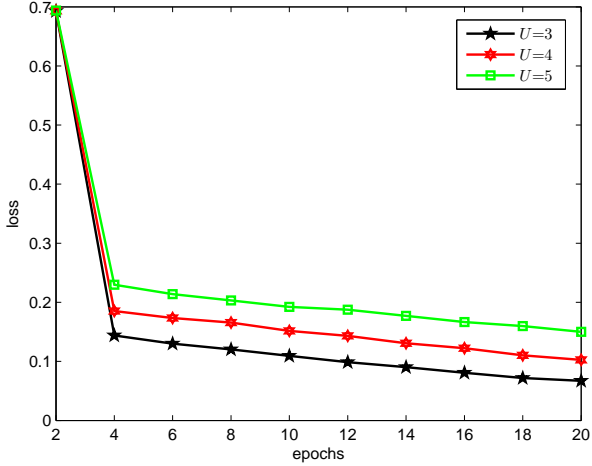}
\caption{Convergence of the training of the NN.}
\label{FIG7}
\end{figure}

\begin{figure}[!t]
\centering
\includegraphics[width=90mm]{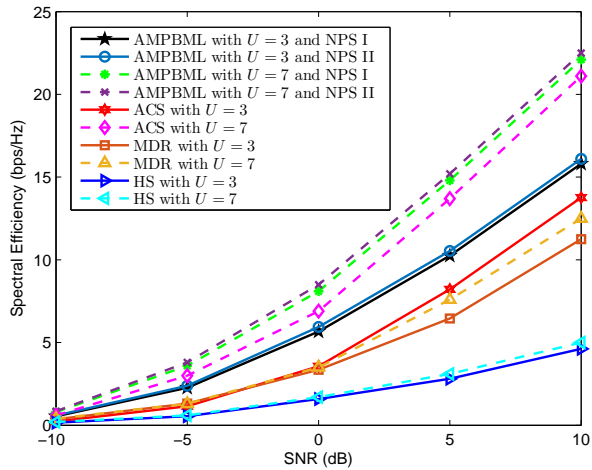}
\caption{Comparisons of spectral efficiency for different SNR.}
\label{FIG4}
\end{figure}

As shown in Fig.~\ref{FIG7}, we verify the convergence of the training of the NN. Suppose there are $U=3,4,5$ users, respectively. It is seen that the loss in \eqref{floss} decreases rapidly as the number of epochs grows, and the less users lead to the smaller loss. For $U=3$, the loss is smaller than 0.1 when the number of epochs is larger than 20, which means 20 epochs is enough to achieve the loss smaller than 0.1.

In Fig.~\ref{FIG4}, we compare the spectral efficiency for different BA methods. During the downlink data transmission, the best analog beamforming vector for the $u$th user is $\boldsymbol{w}(\hat{b}_u)$  . We use $U$ RF chains to design the analog precoder at the BS as $\widetilde{\boldsymbol{W}}_R = \left[ \boldsymbol{w}(\hat{b}_1), \boldsymbol{w}(\hat{b}_2), \ldots, \boldsymbol{w}(\hat{b}_U) \right]$~\cite{sun2019beam}. Then the digital precoder is designed as $\widetilde{\boldsymbol{W}}_B = \left(\widetilde{\boldsymbol{W}}_R^H  \widetilde{\boldsymbol{W}}_R\right)^{-1}$ to eliminate multi-user interference~\cite{sun2019beam}. In order to satisfy the total power constraint, each column of the designed digital precoder, denoted as $\widetilde{\boldsymbol{w}}_{B,u}$, should be normalized, i.e., $\widetilde{\boldsymbol{w}}_{B,u} = \widetilde{\boldsymbol{w}}_{B,u} / \left\| \widetilde{\boldsymbol{w}}_{B,u}^H \widetilde{\boldsymbol{W}}_R^H \right\|_2$, such that $\left\| \widetilde{\boldsymbol{w}}_{B,u}^H \widetilde{\boldsymbol{W}}_R^H \right\| _2 ^2= 1$, $u=1,2,\ldots,U$.  Moreover, since the $U$ beam indices that align with the LOS paths of $U$ users are predicted simultaneously, the user identifier should be transmitted to the BS to determine the one-to-one correspondence of the predicted $U$ beam indices and $U$ users after the hybrid precoder design. Then the spectral efficiency is denoted as~\cite{sun2019beam}
\begin{equation}
  R = \sum_{u=1}^{U} \log_2\left( 1 + \frac{ \frac{1}{U} \left| \widetilde{\boldsymbol{w}}_{B,u}^H \widetilde{\boldsymbol{W}}_R^H \boldsymbol{h}_{u} \right|^2 }{ \frac{1}{U} \sum_{i\neq u} \left| \widetilde{\boldsymbol{w}}_{B,i}^H \widetilde{\boldsymbol{W}}_R^H \boldsymbol{h}_{u} \right|^2 + \sigma^{2} } \right).
\end{equation}
Fig.~\ref{FIG4} shows that the proposed method achieves better performance than the others when $U=3$ and $U=7$.  At SNR = 0 dB, the AMPBML with $U=3$ and NPS I has 59.0\%, 68.7\%, and 257.0\% performance improvement compared with the ACS, MDR, and HS, respectively; while the AMPBML with $U=7$ and NPS I has 17.3\%, 131.4\%, and 376.5\% performance improvement compared with the ACS, MDR, and HS, respectively. The reason for the better performance of  the AMPBML is that it  can train the NN with the received signal containing channel noise and is more effective at low SNR. Moreover, the AMPBML with NPS II has slight performance improvement over that with NPS I, since NPS II uses larger  kernel size and more   filters in each Conv layer.

\section{Conclusions}
In this article, we have proposed an AMPBML  for multi-user mmWave massive MIMO systems. Simulation results have verified the effectiveness of our work. The proposed method can be used in mmWave communication systems to align beams for multiple users simultaneously to reduce the training slots.

\bibliographystyle{IEEEtran}
\bibliography{IEEEabrv,IEEEexample}
\end{document}